\documentclass[10pt,sigconf,letterpaper,authorversion]{acmart}
\usepackage{fancyhdr}
\AtBeginDocument{%
  \providecommand\BibTeX{{%
    \normalfont B\kern-0.5em{\scshape i\kern-0.25em b}\kern-0.8em\TeX}}}

\copyrightyear{2024}
\acmYear{2024}
\setcopyright{acmlicensed}\acmConference[ANRW '24]{Applied Networking Research Workshop}{July 23, 2024}{Vancouver, AA, Canada}
\acmBooktitle{Applied Networking Research Workshop (ANRW '24), July 23, 2024, Vancouver, AA, Canada}
\acmDOI{10.1145/3673422.3674896}
\acmISBN{979-8-4007-0723-0/24/07}

\usepackage{multirow}
\usepackage{booktabs}
\usepackage{todonotes}
\usepackage{graphicx}
\usepackage{subcaption} 
\usepackage{float}
\usepackage{adjustbox}
\usepackage{enumitem}

\begin{document}

\title{To switch or not to switch to TCP Prague? Incentives for adoption
in a partial L4S deployment}

\author{Fatih Berkay Sarpkaya}
\email{fbs6417@nyu.edu}
\affiliation{%
  \institution{New York University}
  \city{Brooklyn}
  \state{NY}
  \country{USA}
}

\author{Ashutosh Srivastava}
\email{as12738@nyu.edu}
\affiliation{%
  \institution{New York University}
  \city{Brooklyn}
  \state{NY}
  \country{USA}
}

\author{Fraida Fund}
\email{ffund@nyu.edu}
\affiliation{%
  \institution{New York University}
  \city{Brooklyn}
  \state{NY}
  \country{USA}
}

\author{Shivendra Panwar}
\email{panwar@nyu.edu}
\affiliation{%
  \institution{New York University}
  \city{Brooklyn}
  \state{NY}
  \country{USA}
}

\begin{abstract}
The Low Latency, Low Loss, Scalable Throughput (L4S) architecture has the potential to reduce queuing delay when it is deployed at endpoints and routers throughout the Internet. However, it is not clear how TCP Prague, a prototype scalable congestion control for L4S, behaves when L4S is not yet universally deployed. 
Specifically, we consider the question: in a partial L4S deployment, will a user benefit by unilaterally switching from the status quo TCP to TCP Prague?
To address this question, we evaluate the performance of a TCP Prague flow when sharing an L4S or non-L4S bottleneck queue with a non-L4S flow.
Our findings suggest that the L4S congestion control, TCP Prague, has less favorable throughput or fairness properties than TCP Cubic or BBR in some coexistence scenarios, which may hinder adoption.
\end{abstract}



\maketitle

\vspace{0.5cm}
 \begin{figure*}
    \centering
    \includegraphics[width=0.95\textwidth]{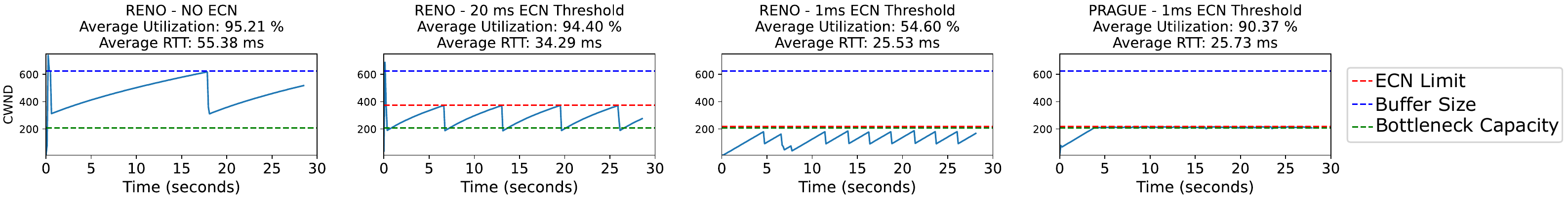}
    \caption{Fundamental Problem of Classic Congestion Control. Adapted from \cite{prague-paper}, we conduct a FABRIC experiment using a line network with 100 Mbps bottleneck capacity, a base RTT of 25 ms, and a buffer size of 2 BDP. The bottleneck AQM is FIFO with ECN. The artifacts to reproduce this experiment are available in \cite{github}.}
    \label{fig: Figure2}
    \vspace{-0.2cm}
\end{figure*}

\section{Introduction}

Low Latency, Low Loss, Scalable Throughput (L4S)~\cite{l4sarch-rfc9330} is an architecture that allows low-delay and classic (non-scalable) data flows to coexist in the same network with reduced latency. It achieves this primarily through three mechanisms: scalable congestion control~\cite{prague-draft-rfc9330,prague-paper}, more accurate Explicit Congestion Notification (AccECN)~\cite{accecn-draft,accecn-rfc7560}, and a dual queue Active Queue Management (AQM)~\cite{dualpi-paper,dualpi-paper2,dualpi-rfc9332}. When all three of these components are in place at the sender, receiver, and bottleneck router, an L4S flow can achieve high throughput with very low latency. However, like any new Internet technology, the deployment of L4S will be incremental. In the initial stages of deployment, L4S flows will coexist with non-L4S flows at L4S or non-L4S bottleneck routers.

This partial L4S deployment scenario is the primary focus of our work. 
In particular, we are interested in the perspective of a sender that has not yet switched to TCP Prague, the scalable congestion control protocol that has been proposed as part of L4S~\cite{prague-draft-rfc9330, prague-paper}.
Given that the bottleneck router may or may not have a dual queue AQM, and given that the other flows sharing the same bottleneck may not be TCP Prague flows, what benefit can a sender expect from unilaterally switching its own congestion control (CC) to TCP Prague? 
This is a key consideration for the eventual deployment of L4S on the Internet. The benefit to individuals deciding whether or not to unilaterally adopt a new technology determines whether or not it will reach a ``tipping point'' and achieve a stable non-zero equilibrium deployment~\cite{networkeffect}.

To address this question, we conduct a series of experiments on the FABRIC~\cite{FABRIC} testbed and measure the throughput and latency of a TCP Prague flow in various partial deployment scenarios.  The results may inform further development of L4S, especially with regard to its behavior in an incremental deployment. While the IETF Transport and Services Working Group (TSVWG) members have been active in evaluating L4S in various scenarios, the academic literature does not cover it extensively. Our work aims to address this gap in the literature, and we will elaborate on this in Sec.~\ref{sec:Background}.

We share our experiment artifacts for an open access testbed (FABRIC)~\cite{FABRIC} so anyone can build on and validate our research\footnote{\label{artifacts}Artifacts are available at: \url{https://github.com/fatihsarpkaya/L4S}}. The rest of this paper is organized as follows. Section~\ref{sec:Background} provides background information on the L4S architecture and its key components. This section also discusses the principles of L4S coexistence and the proposed strategies for its incremental deployment. Section~\ref{sec:Topology} describes the experiment methodology with which we evaluated the coexistence of L4S flows with classic flows and bottleneck routers. Section~\ref{subsec:ExpResults} presents the results of our experiments and offers a detailed analysis of these findings. 

Section~\ref{sec:conclusion} concludes with a summary and directions for future work. 

\section{Background}\label{sec:Background}

The L4S architecture, detailed in \cite{l4sarch-rfc9330}, is designed to reduce network queuing delay using three critical components: scalable congestion control, AccECN, and dual queue AQM. In this section, we describe the key components of the L4S architecture, the state-of-the-art regarding the coexistence of L4S with classic TCP, and the potential hurdles to its deployment from a content provider's perspective.

Classic congestion control, such as TCP Reno or TCP Cubic, responds to network congestion signals such as dropped packets or ECN markings by multiplicatively decreasing its congestion window ($cwnd$), e.g., by a factor of $2$ in TCP Reno. A small ECN threshold would be ideal for maintaining low queuing delays. However, a significant drop in the $cwnd$ at every ECN mark will lead to under-utilization of the link capacity. As depicted in Figure~\ref{fig: Figure2}, for TCP Reno, a high ECN marking threshold prevents under-utilization, but the latency remains high. Setting a low marking threshold (e.g., $1$ ms) causes under-utilization. This highlights a fundamental limitation of traditional ECN-based congestion control mechanisms: the inability to achieve extremely low queuing delays without under-utilizing network capacity.

Scalable congestion controls, like DCTCP, \cite{DCTCP} and TCP Prague, address this issue along with an enhancement to ECN known as AccECN. Using AccECN, a scalable sender can calculate the fraction of ECN-marked packets in the last round and reduce $cwnd$ in proportion to a moving average of this fraction.

The scalable approach tries to react to the extent of congestion and not only its presence. This behavior is also illustrated in Figure~\ref{fig: Figure2}, where TCP Prague achieves high utilization despite a very low ECN threshold.

When a scalable TCP shares a single bottleneck queue (not dual queue or multiple queues) with a classic TCP flow, the difference in their response to ECN marks can cause fairness issues~\cite{dualpi-paper2}. To address this concern, TCP Prague includes an optional ECN fallback heuristic~\cite{prague-fallback} to detect the presence of a single queue, non-L4S, ECN-capable AQM, primarily through RTT variation measurement. On detecting this type of queue, TCP Prague should revert to Reno-like behavior.

While the ECN fallback mechanism prevents TCP Prague from starving classic traffic, we cannot realize the low latency benefits of the L4S solution without upgrading routers. A non-L4S AQM cannot set an extremely low ECN threshold as this is detrimental to classic TCP traffic (Fig.~\ref{fig: Figure2}). Per-flow queuing enables flow isolation, but it would need to enable marking at two different thresholds: a shallow threshold for L4S traffic and a higher threshold for classic traffic. However, as pointed out in~\cite{l4sarch-rfc9330}, per-flow AQMs rely on packet inspection and thus may not be compatible with full end-to-end encryption of transport layer identifiers for privacy and confidentiality, such as IPsec or encrypted VPN tunnels. Per-flow queuing approaches may also not be scalable to core network bottlenecks with thousands of competing flows, a common occurrence in peering links between ISPs~\cite{peering-congestion-1, peering-congestion-2}.

To address this, the Dual-Queue Coupled AQM~\cite{dualpi-rfc9332} separates L4S and non-L4S flows into different queues with different ECN marking thresholds. The marking/drop response of the classic queue is coupled with that of the L4S queue in order to ensure a fair share of available capacity between the classic and low-latency traffic. Hence, the dual queue solution can provide low latency for L4S traffic and achieve fair coexistence without the need for per-flow queuing. The DualPI2 AQM introduced in ~\cite{dualpi-paper} implements this idea.

Like any new Internet protocol, L4S will be deployed in an incremental manner. Since L4S involves changes at TCP senders (scalable congestion control), TCP receivers        (AccECN), and routers (dual queue AQM), a key area of focus in L4S design is the behavior of the protocol when some of these elements are not yet in place. Ideally, to encourage widespread deployment, an L4S flow should have throughput and delay characteristics at least as favorable as a classic flow, even if some elements of the full architecture are missing. Also, an L4S flow should not harm classic flows. 

With respect to incremental deployment, the L4S architecture design~\cite{l4s-ecn-rfc9331} envisions that the L4S AQM will first be deployed on access network bottlenecks in the downstream direction (e.g., as part of low latency DOCSIS~\cite{comcast-docsis}). Then, L4S flows that are part of highly controlled trials will demonstrate the benefit of the architecture. Following this, scalable congestion control and AccECN will be deployed on endpoints to enable more general use of L4S.

However, there are likely to be scenarios in which L4S flows will traverse non-L4S AQMs. First, multiple studies have shown that bottlenecks at peering points between ISPs are also common~\cite{peering-congestion-1, peering-congestion-2}. Also, non-cable access links such as 4G/5G cellular, Wi-Fi, and satellite networks can also be potential bottlenecks. While major wireless network vendors and device manufacturers like Apple are showing interest in L4S~\cite{l4s-interop}, widespread deployment is still some time away. Upgrading legacy access network routers, e.g., Wi-Fi routers or 3G/4G base stations worldwide, to support L4S will also be a challenge. Thus, the behavior of an L4S flow over a non-L4S AQM is of great interest to content providers who might consider switching to scalable congestion control.

Field trials and prototype demonstrations validate the low latency benefits of the L4S architecture in controlled environments~\cite{nokia-hololight, ietf-l4s-deployment}.
However, some early work evaluating L4S~\cite{sce-l4s-bakeoff,l4s-tests,henderson2019l4s-issues}, has raised concerns about partial deployment. Although most of this work has not been published in the academic literature, these results have been used in IETF meetings and discussions~\cite{IETF-112,interim-2020-issue-16} and to inform further protocol development.
It is shown in \cite{l4s-tests} and validated in ~\cite{henderson2019l4s-issues} that L4S flows dominate both Cubic and Reno flows in a single-queue bottleneck with ECN. The ECN Fallback heuristic~\cite{prague-fallback} has been proposed to address this issue, but previous work on low latency congestion control suggests that the delay variance measure used in this heuristic is not a reliable signal in all settings, and that in fact, there is no universally effective signal of sharing a bottleneck with classic flow~\cite{srivastava2022coexistence, budzisz2011fair, ecn_or_delay}.
Additionally, \cite{l4s-tests} mentions that DualPI2 consistently provides a throughput advantage to Cubic flows, and~\cite{boruoljira2020validating} finds that the fairness of the DualPI2 AQM is not robust to small variations in protocol implementations. 

In this work, we seek to validate these early findings, evaluate the effectiveness of the ECN Fallback heuristic,  and to also consider scenarios where a TCP Prague flow shares a bottleneck with a TCP BBR flow (as a substantial portion of Internet traffic uses BBR~\cite{census}).

\section{Experiment Methodology}\label{sec:Topology}

In this section, we describe our experiment setup for evaluating the performance of TCP Prague when sharing a bottleneck link with TCP Cubic or TCP BBR.

\textbf{Experiment Platform:} We conduct experiments on FABRIC \cite{FABRIC}, a national scale programmable experimental networking testbed. Each node in our experiment is a virtual machine running Ubuntu 22.04, with 4 cores and 32 GB RAM.

\begin{figure}
    \centering    \includegraphics[width=0.31\textwidth]{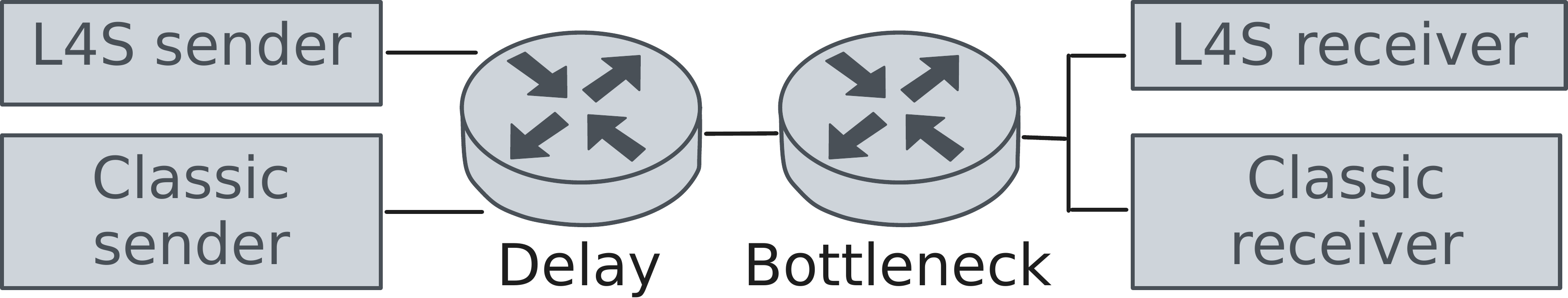}
    \caption{Experiment Topology}
    \label{fig: Figure4}
    \vspace{-0.4cm}
\end{figure}

\textbf{Topology:} We use a line topology comprising two senders and two receivers (L4S and classic), connected via a delay node and a bottleneck router, as illustrated in Figure~\ref{fig: Figure4}. At the delay node, we use \texttt{netem} to emulate a base RTT, with half of the delay added in each direction. At the bottleneck router, we use the token bucket filter implemented in \texttt{tc-htb} to configure the bottleneck bandwidth and buffer size.

\textbf{Network Settings:} On the Internet, a flow is most likely to encounter a bottleneck either at a peering point, or at the access link. For these experiments, we emulate network conditions that are representative of an access link: 10~ms base RTT, 100~Mbps bottleneck link capacity. 

\textbf{Queue:} Our experiments consider the following queue configurations at the bottleneck router:
\begin{itemize}[noitemsep,topsep=0pt]
  \item \textbf{FIFO:} a single drop tail queue without ECN support, realized with \texttt{tc-bfifo}.

  \item \textbf{FIFO + ECN:} a single drop tail queue with ECN support using a static 5~ms marking threshold, realized with \texttt{tc-fq}. (Although \texttt{tc-fq} is multi-queue, by setting the \texttt{orphan mask} option to 0, we enforce that all flows are hashed to a single queue.)
  
  \item \textbf{CoDel:} a single queue with CoDel AQM~\cite{codel}, which uses the local minimum queue size within a monitoring window as a measure of the standing queue, and marks packets if there is a standing queue exceeding a target value. We use \texttt{tc-codel} with a 5~ms target and the ECN option enabled, so that it marks packets for flows with ECN support and drops packets otherwise.
  
  \item \textbf{FQ:} a fair queue with flow isolation and ECN (using a static 5~ms marking threshold), realized with \texttt{tc-fq}. 

  \item \textbf{FQ-CoDel:} combines fair queuing with the CoDel AQM. We realize this queue with \texttt{tc fq\_codel}, with a 5~ms target and the ECN option enabled. 
  
  \item \textbf{DualPI2:} a dual queue coupled AQM designed for L4S \cite{dualpi-rfc9332}, realized using \texttt{tc-dualpi2} from the L4S repository \cite{l4srepo} (commit \texttt{4579ffb}). The \texttt{target} parameter for the Proportional Integral (PI) controller is 5~ms and the \texttt{step.thresh} parameter for the L4S queue is 1~ms .
  
\end{itemize}

For each type of queue, we consider bottleneck buffer sizes that are shallow and deep, including the following multiples of the link bandwidth delay product (BDP): 0.5, 1, 2, 4, and 8.

\textbf{Congestion control and AccECN at endpoints:}  In our experiments, L4S flows run TCP Prague, using the implementation in the L4S repository \cite{l4srepo} (commit \texttt{4579ffb}). The L4S sender and receiver support AccECN. We also evaluate TCP Prague with and without the ECN Fallback heuristic~\cite{prague-fallback}, which is an optional setting that needs to be turned on explicitly. For the classic flow, we consider two CCs that a TCP Prague flow is likely to encounter at a shared bottleneck:\\
- \textbf{TCP Cubic} \cite{cubic} - according to a recent estimate, this is the predominant TCP variant on the Internet~\cite{census}. We consider a classic Cubic flow (implementation in Linux kernel 5.13.12) with and without ECN support. \\
- \textbf{TCP BBR} - BBR and its variants account for 22\% of the top websites and approximately 40\% of Internet traffic by volume~\cite{census}. We conduct experiments with BBRv1 (implementation in Linux kernel 5.13.12), which does not support ECN. We also consider BBRv2, with and without ECN, using the implementation in the \texttt{v2alpha} branch of the official BBR repository~\cite{bbrrepo}. 
Finally, we also run experiments with an L4S-compatible BBRv2 flow, using the BBRv2 with AccECN support from the L4S repository~\cite{l4srepo} (commit \texttt{4579ffb}).

\textbf{Flow Generation}: For each network configuration, we generate a single TCP Prague flow from the L4S sender and a single TCP Cubic or TCP BBRv1/v2 flow from the classic sender using the \texttt{iperf3} utility, for a duration of 60 seconds. We then record the average throughput and RTT values for each flow. The results presented are the average of 10 trials.

\vspace{-0.5cm}
\section{Experiment Results}\label{subsec:ExpResults}

\begin{figure*}
  \centering
  \includegraphics[width=0.86\textwidth]{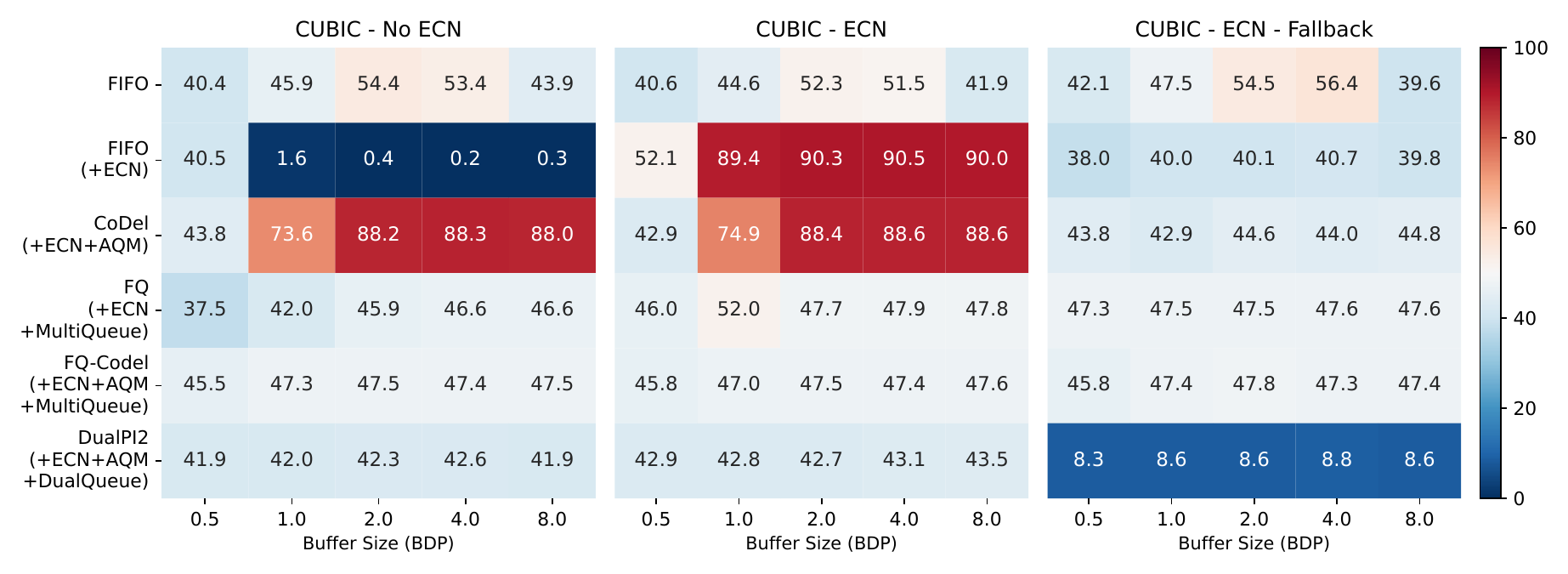}
  \vspace{-0.2cm}
  \caption{Prague throughput (Mbps) when sharing 100 Mbps bottleneck with Cubic flow.}
  \label{fig:Prague Throughput vs Cubic}

  \centering
  \includegraphics[width=0.86\textwidth]{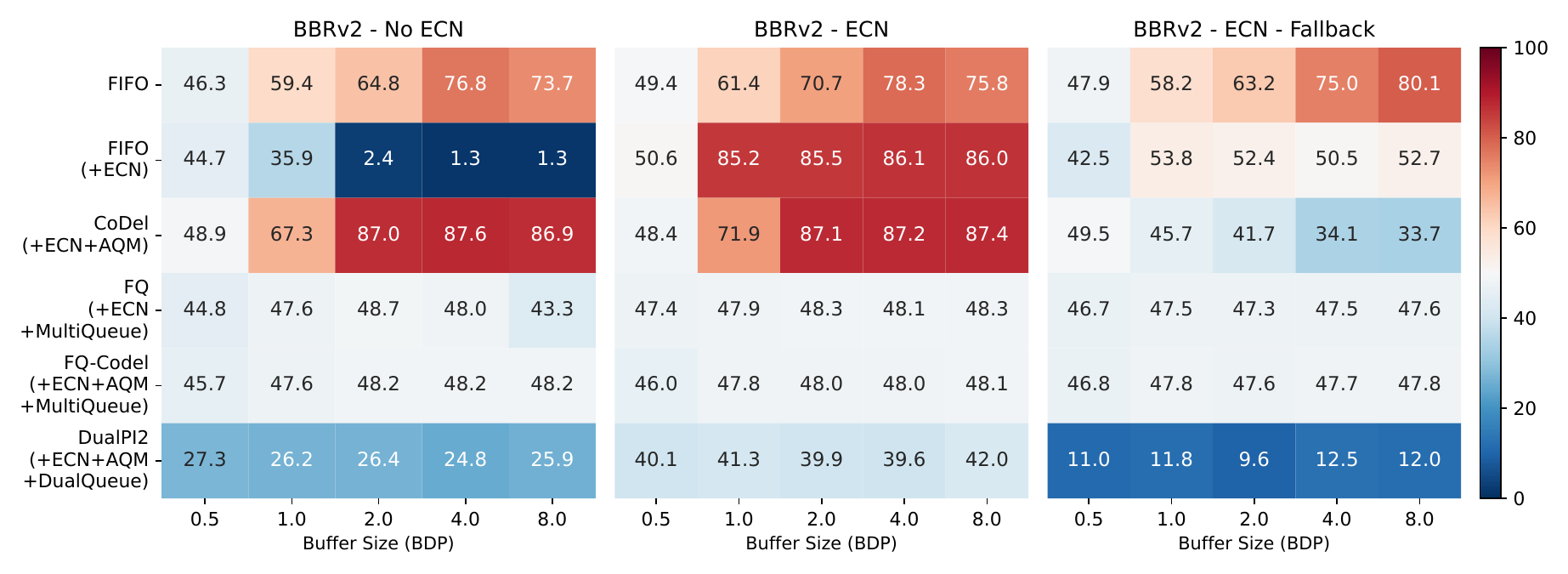}
  \vspace{-0.2cm}
  \caption{Prague throughput (Mbps) when sharing 100 Mbps bottleneck with BBRv2 flow.}
  \label{fig:Prague Throughput vs BBRv2}

  \centering
  \includegraphics[width=0.86\textwidth]{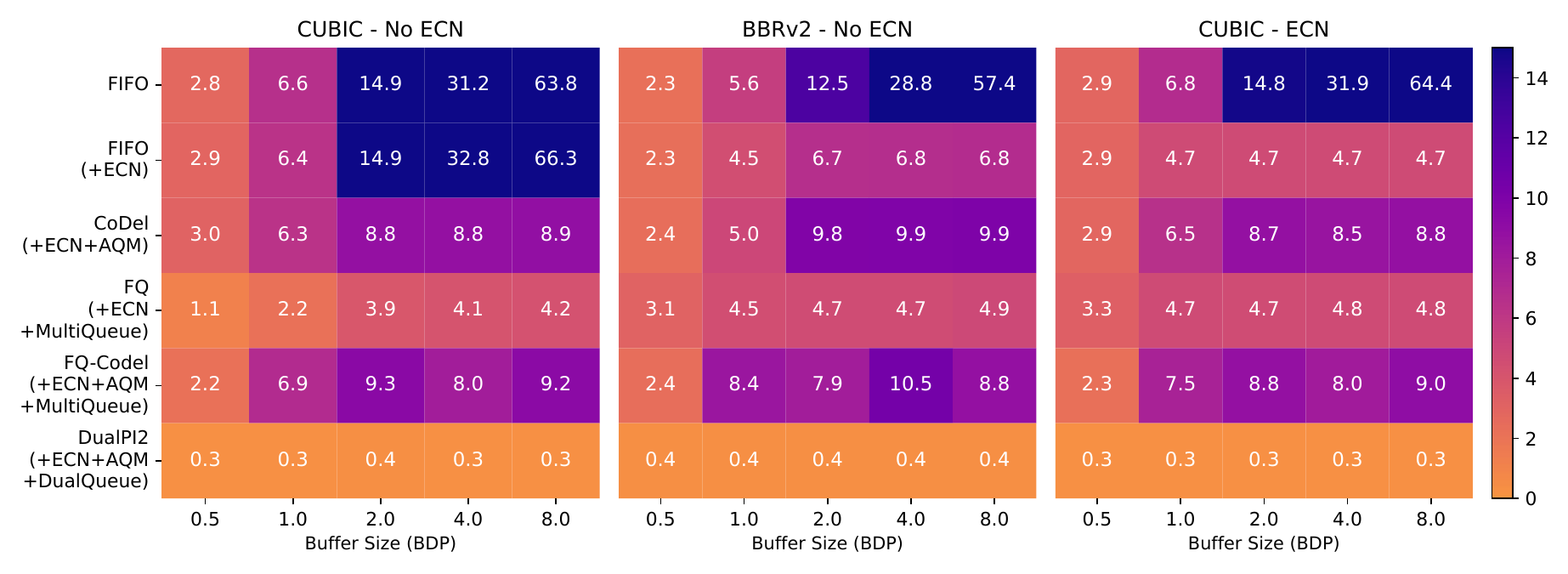}
  \vspace{-0.2cm}
  \caption{Prague queuing delay (ms) when sharing bottleneck with classic flow. (ECN threshold is 5 ms, where applicable. For DualPI2, L4S queue has 1 ms threshold.)}
  \label{fig: Prague Latency}
\end{figure*}

\begin{figure}[t]
  \centering
  \begin{subfigure}{.4\textwidth}
    \includegraphics[width=\textwidth]{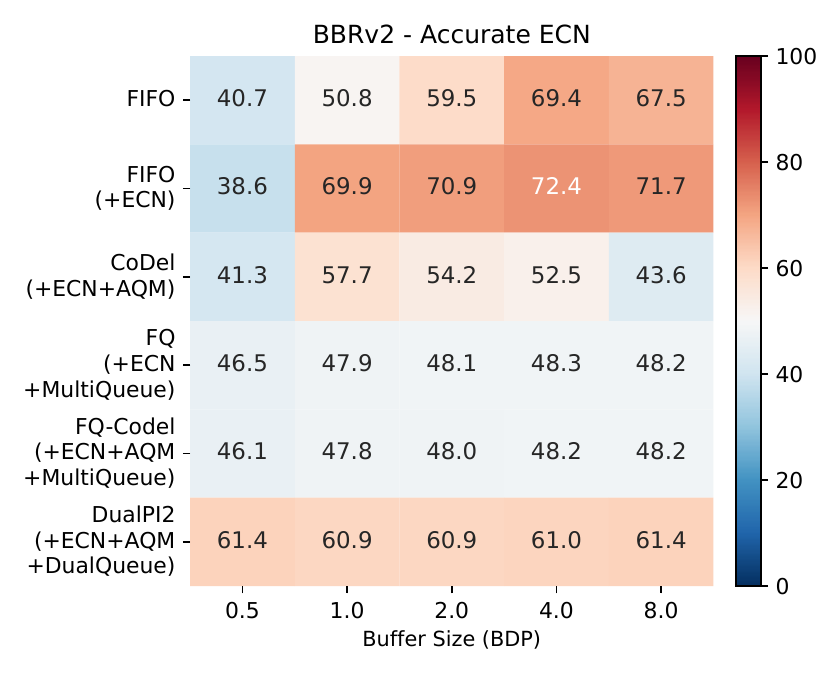}
      \vspace{-0.5cm}
    \caption{Prague throughput (Mbps)}
    \label{fig:tput-bbrv2-accecn}
  \end{subfigure}
  \\
  \begin{subfigure}{.4\textwidth}
    \includegraphics[width=\textwidth]{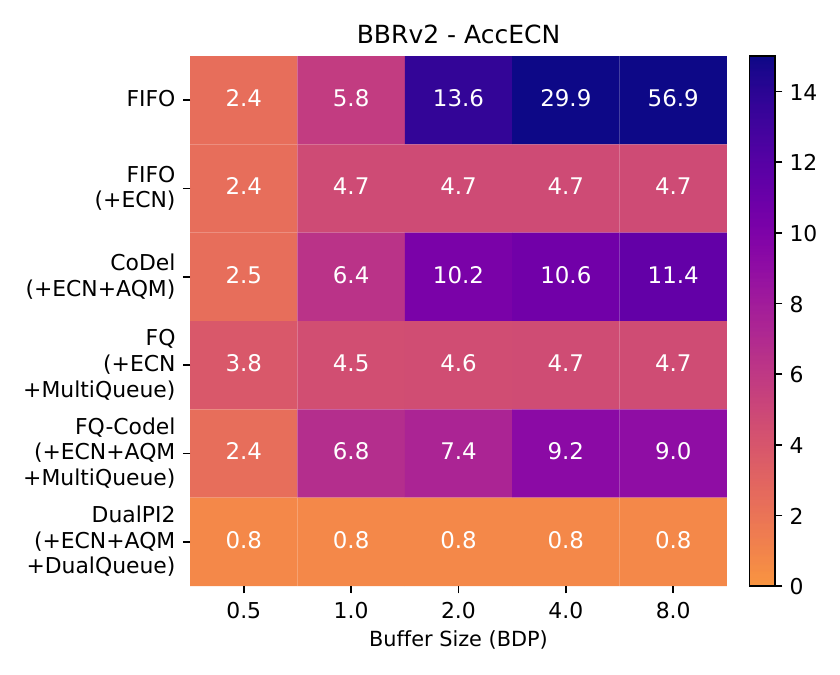}
  \vspace{-0.5cm}
    \caption{Prague queuing delay (ms)}
    \label{fig:latency-bbrv2-accecn}
  \end{subfigure}
  \vspace{-0.2cm}
  \caption{Prague throughput and queuing delay, sharing a 100 Mbps bottleneck with an AccECN BBRv2 flow.}
  \label{fig:bbrv2-accecn}
  \vspace{-0.6cm}
\end{figure}

In this section, we present and evaluate the throughput and queuing latency of a TCP Prague flow when it shares a single bottleneck with a competing flow (Cubic or BBR) under various previously described network conditions. The main question we would like to answer is: \textbf{should content originators use L4S (TCP Prague) as their congestion control protocol?} Our results are presented in Figures~\ref{fig:Prague Throughput vs Cubic},~\ref{fig:Prague Throughput vs BBRv2},~\ref{fig: Prague Latency} and \ref{fig:bbrv2-accecn}. Here, we discuss some of the major findings.

\textbf{Prague throughput is degraded} when sharing a single, ECN-enabled queue without AQM alongside a BBRv2 or Cubic flow that \textit{does not respond to ECN} signals. This is shown in Fig.~\ref{fig:Prague Throughput vs Cubic} (Cubic - No ECN) and Fig.~\ref{fig:Prague Throughput vs BBRv2} (BBRv2 - No ECN) for the \textbf{FIFO + ECN} AQM. This outcome is expected because if the classic sender does not respond to ECN, it fills the bottleneck buffer. However, we want to highlight this as a concern for L4S deployment because such a scenario can occur when an endpoint does not support ECN or an Internet path encounters ``ECN bleaching'', i.e., an intermediate network device clearing the ECN flags. Both of these remain problematic according to a recent measurement study~\cite{lim-ecn-traversal}. 

Although not shown here, the same trend is observed when competing with a BBRv1 flow\footnotemark[\value{footnote}]. Prague is starved by BBRv1 in shallow buffer settings for most non-L4S queues (except FQ-Codel). BBRv1's domination over TCP Cubic in shallow buffers has also been observed in previous work~\cite{bbr-2019-ware}.

We observe that Prague is dominated by a BBRv2 flow whose endpoints do not support ECN when sharing an L4S AQM queue (\textbf{Fig.~\ref{fig:Prague Throughput vs BBRv2}, DualPI2}). The DualPI2 coupling design assumes the classic queue carries loss-based TCP traffic (Cubic or Reno) but the same coupling parameters do not work with BBR flows in the classic queue. This raises a broader concern that the dual queue coupled AQM strategy may not generalize well when an L4S CC competes with non-L4S traffic that is not loss-based TCP. We further observe that even when competing with Cubic, Prague achieves slightly lower throughput (around 40\%) under DualPI2 AQM regardless of ECN support, contradicting the observation in \cite{l4s-tests} that DualPI2 provides a throughput advantage to Prague flows.

\textbf{Prague takes more than its fair share of throughput} when competing with a Cubic or BBRv2 flow that \textit{responds to ECN} while sharing a single, ECN-enabled queue (FIFO $+$ ECN or Codel $+$ ECN). This arises from differences in TCP Prague's and other TCP's response to ECN signals, as explained in section \ref{sec:Background}. Our results validate the observations in \cite{l4s-tests} for single queue with classic ECN AQMs. This scenario motivated the implementation of the ECN fallback heuristic in TCP Prague, which we also evaluated. With ECN fallback turned ON for TCP Prague, we see improved coexistence between Prague and ECN-capable classic flows over single queue ECN AQMs. However, we observed other problems (to be discussed shortly) with ECN fallback.

An AQM like CoDel drops packets from classic flows that \textit{do not react to ECN} to maintain queuing delay close to its target. Prague again dominates the classic flows in this case (\textbf{Fig.~\ref{fig:Prague Throughput vs Cubic} \& ~\ref{fig:Prague Throughput vs BBRv2}, CoDel}) because of its scalable \texttt{cwnd} drop strategy. 

\textbf{When there is a non-ECN bottleneck,} there is no ECN signal, and Prague falls back to operating like TCP Reno to maintain friendliness to classic TCP traffic. In our experiment conditions, Prague then achieves a fair share of throughput when competing with TCP Cubic (\textbf{Fig.~\ref{fig:Prague Throughput vs Cubic}, FIFO}). This finding disagrees with \cite{l4s-tests} where Cubic flows dominate Prague flows in FIFO bottlenecks without ECN. However, the buffer size used in this work was much higher than the maximum buffer of $8$ BDP used in our experiments. 

On the other hand, Prague dominates the BBRv2 flow over a FIFO queue (\textbf{Fig.~\ref{fig:Prague Throughput vs BBRv2}, FIFO}). A similar result was observed for BBRv1 in our experiments and is a well-documented issue for BBR's coexistence with loss-based TCP in deep buffers~\cite{bbr-2019-ware}. 

\textbf{Prague gets its fair share of throughput} when sharing an FQ bottleneck, regardless of whether the competing flow is TCP Cubic or BBRv2, or whether the Fallback algorithm is ON or OFF (\textbf{Fig.~\ref{fig:Prague Throughput vs Cubic} \&~\ref{fig:Prague Throughput vs BBRv2}, FQ + ECN \& FQ-CoDel + ECN}). This confirms the claim from \cite{l4s-tests} that the CodelAF AQM allows for fair sharing between Cubic and Prague flows, as verified by our results with per-flow queuing (FQ) bottlenecks.

\textbf{For latency-sensitive applications,} the only case where we observe TCP Prague consistently getting sub-$1$ ms queuing delays is with an L4S AQM  that can mark L4S packets at a shallow threshold (like DualPI2) (\textbf{Fig.~\ref{fig: Prague Latency}}). 
The ECN fallback heuristic, aimed at preventing Prague from dominating classic traffic in classic AQMs, may not benefit content originators. It may detect a classic queue even in cases where the bottleneck is DualPI2.  When this happens, TCP Prague uses classic TCP behavior, and is starved because of the $1$ ms threshold of the L4S queue (\textbf{Fig.~\ref{fig:Prague Throughput vs Cubic}, \ref{fig:Prague Throughput vs BBRv2}, Fallback, DualPI2}). 

\textbf{If the BBRv2 flow supports accurate ECN,} it can also experience the low latency benefits of L4S. With a DualPI2 bottleneck, both the Prague and AccECN enabled BBRv2 are classified into the low latency queue, leading to sub-$1$ ms queuing delays (\textbf{Fig.~\ref{fig:latency-bbrv2-accecn}}). The throughput share is $60:40$ in Prague's favor (\textbf{Fig.~\ref{fig:tput-bbrv2-accecn}}).

\vspace{1cm}
\begin{table}[t]
\begin{tabular}{|c|cc|ll|}
\hline                                                                  
\multirow{2}{*}{Buffer Type}                   & \multicolumn{2}{c|}{ECN Fallback OFF}                   & \multicolumn{2}{c|}{ECN Fallback ON}                    \\ \cline{2-5} 
                                               & \multicolumn{1}{c|}{Cubic} & \multicolumn{1}{c|}{BBRv2} & \multicolumn{1}{c|}{Cubic} & \multicolumn{1}{c|}{BBRv2} \\ \hline
\multicolumn{1}{|l|}{SQ w/o ECN} & \multicolumn{1}{c|}{\checkmark}      & \multicolumn{1}{c|}{X}     & \multicolumn{1}{c|}{\checkmark}      & \multicolumn{1}{c|}{X}     \\ \hline
SQ + ECN                             & \multicolumn{1}{c|}{X}     & \multicolumn{1}{c|}{X}     & \multicolumn{1}{c|}{\checkmark}      &  \multicolumn{1}{c|}{\checkmark}                          \\ \hline
FQ + ECN                             & \multicolumn{1}{c|}{\checkmark}      & \multicolumn{1}{c|}{\checkmark}      & \multicolumn{1}{c|}{\checkmark}      &  \multicolumn{1}{c|}{\checkmark}                          \\ \hline
DualPI2                                        & \multicolumn{1}{c|}{\checkmark}      & \multicolumn{1}{c|}{\checkmark}      & \multicolumn{1}{c|}{X}     & \multicolumn{1}{c|}{X}     \\ \hline
\end{tabular}
\caption{Is it okay to turn on TCP Prague (\checkmark) or not (X)? \\(SQ: single queue, FQ: fair queuing)} 
\label{Table 1}
\vspace{-0.8cm}
\end{table}

\vspace{-0.9cm}
Table~\ref{Table 1} summarizes our findings into recommendations on using the TCP Prague CC. Since the TCP sender does not know what type of bottleneck it will encounter, the decision of whether or not to turn on TCP Prague at the sender depends on the expected types of queues likely to be deployed at bottlenecks and the types of flows likely to traverse these queues. If the bottleneck uses fair queuing (FQ), TCP Prague can be safely turned ON. While the ECN fallback algorithm improves fairness in single queue bottlenecks, it is prone to misdetection of dual queue AQMs, which can lead to Prague being starved even when AQMs compatible with the L4S architecture are deployed, e.g., DualPI2. This finding motivates the need to engineer better solutions for TCP Prague coexistence in non-L4S bottlenecks (for encouraging incremental deployment), to the extent possible~\cite{arun2022starvation}, or at least to ensure that Prague will not be harmful~\cite{ware-jain-fairness}. Additionally, it is important that these solutions do not degrade Prague's performance in an L4S-enabled network.

\section{Conclusion}
\label{sec:conclusion}

In this work, we have investigated the conditions under which switching to TCP Prague is beneficial for a content originator and/or safe for other flows. Our findings suggest that if the content originator of TCP Prague cannot be sure what type of queue is at the bottleneck router, a range of outcomes for throughput and latency are possible, some of which are unfavorable. It is demonstrated that bottlenecks with per-flow isolation are the only type that ensures fairness, while dual queue AQM bottlenecks are the only type that guarantees low latency. In certain scenarios, TCP Prague may consume much more than its fair share of the link capacity when competing with Cubic, BBRv1, or BBRv2, and in other scenarios, TCP Prague gets much less than its fair share. 

For future work, we hope to extend this analysis to more diverse network environments, including multiple flows, multiple bottlenecks, and more realistic traffic patterns. We will also consider other scalable congestion control algorithms that may be relevant to a content provider adopting L4S.

\begin{acks}
This research was supported by the New York State Center for Advanced Technology in Telecommunications and Distributed Systems (CATT), NYU WIRELESS, and the National Science Foundation (NSF) under Grant No. CNS-2148309 and OAC-2226408.
\end{acks}
\newpage
\bibliographystyle{IEEEtran}
\bibliography{ref.bib}

\begin{thebibliography}{10}
\providecommand{\url}[1]{#1}
\csname url@samestyle\endcsname
\providecommand{\newblock}{\relax}
\providecommand{\bibinfo}[2]{#2}
\providecommand{\BIBentrySTDinterwordspacing}{\spaceskip=0pt\relax}
\providecommand{\BIBentryALTinterwordstretchfactor}{4}
\providecommand{\BIBentryALTinterwordspacing}{\spaceskip=\fontdimen2\font plus
\BIBentryALTinterwordstretchfactor\fontdimen3\font minus \fontdimen4\font\relax}
\providecommand{\BIBforeignlanguage}[2]{{%
\expandafter\ifx\csname l@#1\endcsname\relax
\typeout{** WARNING: IEEEtran.bst: No hyphenation pattern has been}%
\typeout{** loaded for the language `#1'. Using the pattern for}%
\typeout{** the default language instead.}%
\else
\language=\csname l@#1\endcsname
\fi
#2}}
\providecommand{\BIBdecl}{\relax}
\BIBdecl

\bibitem{prague-paper}
B.~Briscoe, K.~De~Schepper, O.~Tilmans, M.~K{\"u}hlewind, J.~Misund, O.~Albisser, and A.~S. Ahmed, ``{Implementing the ’Prague Requirements’ for Low Latency Low Loss Scalable Throughput (L4S)},'' \emph{Netdev 0x13}, 2019.

\bibitem{github}
F.~F. Fatih Berkay~Sarpkaya, ``{Reproducing "Scalable Congestion Control Resolves the Delay Utilization Dilemma"},'' \url{https://github.com/fatihsarpkaya/TCP-ECN}, 2024.

\bibitem{l4sarch-rfc9330}
\BIBentryALTinterwordspacing
B.~Briscoe, K.~D. Schepper, M.~Bagnulo, and G.~White, ``{Low Latency, Low Loss, and Scalable Throughput (L4S) Internet Service: Architecture},'' RFC 9330, Jan. 2023. [Online]. Available: \url{https://datatracker.ietf.org/doc/html/rfc9330}
\BIBentrySTDinterwordspacing

\bibitem{prague-draft-rfc9330}
\BIBentryALTinterwordspacing
K.~D. Schepper, O.~Tilmans, B.~Briscoe, and V.~Goel, ``{Prague Congestion Control},'' Internet Engineering Task Force, Internet-Draft draft-briscoe-iccrg-prague-congestion-control-03, Oct. 2023, work in Progress. [Online]. Available: \url{https://datatracker.ietf.org/doc/draft-briscoe-iccrg-prague-congestion-control/03/}
\BIBentrySTDinterwordspacing

\bibitem{accecn-draft}
\BIBentryALTinterwordspacing
B.~Briscoe, M.~Kühlewind, and R.~Scheffenegger, ``{More Accurate Explicit Congestion Notification (ECN) Feedback in TCP},'' Internet Engineering Task Force, Internet-Draft draft-ietf-tcpm-accurate-ecn-28, Nov. 2023, work in Progress. [Online]. Available: \url{https://datatracker.ietf.org/doc/draft-ietf-tcpm-accurate-ecn/28/}
\BIBentrySTDinterwordspacing

\bibitem{accecn-rfc7560}
\BIBentryALTinterwordspacing
M.~Kühlewind, R.~Scheffenegger, and B.~Briscoe, ``{Problem Statement and Requirements for Increased Accuracy in Explicit Congestion Notification (ECN) Feedback},'' RFC 7560, Aug. 2015. [Online]. Available: \url{https://datatracker.ietf.org/doc/html/rfc7560}
\BIBentrySTDinterwordspacing

\bibitem{dualpi-paper}
O.~Albisser, K.~De~Schepper, B.~Briscoe, O.~Tilmans, and H.~Steen, ``{DUALPI2—Low Latency, Low Loss and Scalable (L4S) AQM},'' NetDev 0x13, Prague, 2019.

\bibitem{dualpi-paper2}
K.~D. Schepper, O.~Albisser, O.~Tilmans, and B.~Briscoe, ``{Dual Queue Coupled AQM: Deployable Very Low Queuing Delay for All},'' 2022.

\bibitem{dualpi-rfc9332}
\BIBentryALTinterwordspacing
K.~D. Schepper, B.~Briscoe, and G.~White, ``{Dual-Queue Coupled Active Queue Management (AQM) for Low Latency, Low Loss, and Scalable Throughput (L4S)},'' RFC 9332, Jan. 2023. [Online]. Available: \url{https://datatracker.ietf.org/doc/html/rfc9332}
\BIBentrySTDinterwordspacing

\bibitem{networkeffect}
N.~Economides, ``{The Economics of Networks},'' \emph{Intl. Journal of Industrial Organization}, vol.~14, no.~6, pp. 673--699, 1996.

\bibitem{FABRIC}
I.~Baldin, A.~Nikolich, J.~Griffioen, I.~I.~S. Monga, K.-C. Wang, T.~Lehman, and P.~Ruth, ``{FABRIC: A National-Scale Programmable Experimental Network Infrastructure},'' \emph{IEEE Internet Computing}, vol.~23, no.~6, pp. 38--47, 2019.

\bibitem{DCTCP}
\BIBentryALTinterwordspacing
M.~Alizadeh, A.~Greenberg, D.~A. Maltz, J.~Padhye, P.~Patel, B.~Prabhakar, S.~Sengupta, and M.~Sridharan, ``{Data Center TCP (DCTCP)},'' \emph{SIGCOMM Comput. Commun. Rev.}, vol.~40, no.~4, p. 63–74, aug 2010. [Online]. Available: \url{https://doi.org/10.1145/1851275.1851192}
\BIBentrySTDinterwordspacing

\bibitem{prague-fallback}
\BIBentryALTinterwordspacing
B.~Briscoe and A.~S. Ahmed, ``{TCP Prague Fall-back on Detection of a Classic ECN AQM},'' 2021. [Online]. Available: \url{https://arxiv.org/abs/1911.00710}
\BIBentrySTDinterwordspacing

\bibitem{peering-congestion-1}
\BIBentryALTinterwordspacing
A.~Dhamdhere, D.~D. Clark, A.~Gamero-Garrido, M.~Luckie, R.~K.~P. Mok, G.~Akiwate, K.~Gogia, V.~Bajpai, A.~C. Snoeren, and K.~Claffy, ``{Inferring persistent interdomain congestion},'' in \emph{{Proceedings of the 2018 Conference of the ACM Special Interest Group on Data Communication}}, ser. SIGCOMM '18.\hskip 1em plus 0.5em minus 0.4em\relax New York, NY, USA: Association for Computing Machinery, 2018, p. 1–15. [Online]. Available: \url{https://doi.org/10.1145/3230543.3230549}
\BIBentrySTDinterwordspacing

\bibitem{peering-congestion-2}
\BIBentryALTinterwordspacing
A.~Akella, S.~Seshan, and A.~Shaikh, ``{An empirical evaluation of wide-area internet bottlenecks},'' in \emph{{Proceedings of the 2003 ACM SIGMETRICS International Conference on Measurement and Modeling of Computer Systems}}, ser. SIGMETRICS '03.\hskip 1em plus 0.5em minus 0.4em\relax New York, NY, USA: Association for Computing Machinery, 2003, p. 316–317. [Online]. Available: \url{https://doi.org/10.1145/781027.781075}
\BIBentrySTDinterwordspacing

\bibitem{l4s-ecn-rfc9331}
\BIBentryALTinterwordspacing
K.~D. Schepper and B.~Briscoe, ``{The Explicit Congestion Notification (ECN) Protocol for Low Latency, Low Loss, and Scalable Throughput (L4S)},'' RFC 9331, Jan. 2023. [Online]. Available: \url{https://datatracker.ietf.org/doc/rfc9331/}
\BIBentrySTDinterwordspacing

\bibitem{comcast-docsis}
J.~Livingood, ``{Comcast Kicks Off Industry’s First Low Latency DOCSIS Field Trials},'' \url{https://corporate.comcast.com/press/releases/comcast-multi-gig-symmetrical-speeds-world-first-docsis-4-deployment}, 2023.

\bibitem{l4s-interop}
G.~White, ``{L4S Interop Lays Groundwork for 10G Metaverse},'' \url{https://www.cablelabs.com/blog/l4s-interop-lays-groundwork-for-10g-metaverse}, 2022.

\bibitem{nokia-hololight}
N.~Corporation, ``{Nokia collaborates with Hololight to deliver reliable immersive XR experiences with latency-improving technology L4S},'' \url{https://www.nokia.com/about-us/news/releases/2023/11/02/nokia-collaborates-with-hololight-to-deliver-reliable-immersive-xr-experiences-with-latency-improving-technology-l4s/}, 2023.

\bibitem{ietf-l4s-deployment}
J.~Livingood, ``{Comcast L4S Field Trial Update},'' \url{https://datatracker.ietf.org/meeting/118/materials/slides-118-tsvwg-sessa-61-l4s-experience-01}, 2023.

\bibitem{sce-l4s-bakeoff}
P.~Heist, ``{sce-l4s-bakeoff},'' \url{https://github.com/heistp/sce-l4s-bakeoff}, 2019.

\bibitem{l4s-tests}
------, ``{L4S Tests},'' \url{https://github.com/heistp/l4s-tests}, 2021.

\bibitem{henderson2019l4s-issues}
\BIBentryALTinterwordspacing
T.~Henderson, O.~Tilmans, and G.~White, ``{Testbed and Simulation Results for TSVWG Scenarios},'' 2019, accessed: 2024-06-12. [Online]. Available: \url{https://l4s.cablelabs.com/l4s_issues.html}
\BIBentrySTDinterwordspacing

\bibitem{IETF-112}
\BIBentryALTinterwordspacing
G.~W. Bob~Briscoe, Koen De~Schepper, ``{L4S Status Update},'' Presented at IETF 112, Online, 2021, accessed: 2024-06-12. [Online]. Available: \url{https://datatracker.ietf.org/meeting/112/materials/slides-112-tsvwg-sessa-32-l4s-ecn-drafts-01.pdf}
\BIBentrySTDinterwordspacing

\bibitem{interim-2020-issue-16}
O.~T. G.~W. Bob~Briscoe, Koen De~Schepper, ``{Low Latency Low Loss Scalable Throughput (L4S)},'' \url{https://www.ietf.org/proceedings/interim-2020-tsvwg-01/slides/slides-interim-2020-tsvwg-01-sessa-l4s-tcp-prague-update-00.pdf}, February 2020, interim 2020 TSVWG Meeting.

\bibitem{srivastava2022coexistence}
A.~Srivastava, F.~Fund, and S.~S. Panwar, ``{Coexistence of delay-based TCP congestion control: Challenges and opportunities},'' in \emph{{2022 IEEE International Workshop Technical Committee on Communications Quality and Reliability (CQR)}}.\hskip 1em plus 0.5em minus 0.4em\relax IEEE, 2022, pp. 43--48.

\bibitem{budzisz2011fair}
{\L}.~Budzisz, R.~Stanojevic, A.~Schlote, F.~Baker, and R.~Shorten, ``{On the fair coexistence of loss-and delay-based TCP},'' \emph{IEEE/ACM transactions on networking}, vol.~19, no.~6, pp. 1811--1824, 2011.

\bibitem{ecn_or_delay}
\BIBentryALTinterwordspacing
Y.~Zhu, M.~Ghobadi, V.~Misra, and J.~Padhye, ``{ECN or Delay: Lessons Learnt from Analysis of DCQCN and TIMELY},'' in \emph{{Proceedings of the 12th International on Conference on Emerging Networking EXperiments and Technologies}}, ser. CoNEXT '16.\hskip 1em plus 0.5em minus 0.4em\relax New York, NY, USA: Association for Computing Machinery, 2016, p. 313–327. [Online]. Available: \url{https://doi.org/10.1145/2999572.2999593}
\BIBentrySTDinterwordspacing

\bibitem{boruoljira2020validating}
D.~BoruOljira, K.-J. Grinnemo, A.~Brunstrom, and J.~Taheri, ``{Validating the sharing behavior and latency characteristics of the L4S architecture},'' \emph{ACM SIGCOMM Computer Communication Review}, vol.~50, no.~2, pp. 37--44, 2020.

\bibitem{census}
\BIBentryALTinterwordspacing
A.~Mishra, X.~Sun, A.~Jain, S.~Pande, R.~Joshi, and B.~Leong, ``{The Great Internet TCP Congestion Control Census},'' \emph{Proc. ACM Meas. Anal. Comput. Syst.}, vol.~3, no.~3, dec 2019. [Online]. Available: \url{https://doi.org/10.1145/3366693}
\BIBentrySTDinterwordspacing

\bibitem{codel}
\BIBentryALTinterwordspacing
K.~Nichols and V.~Jacobson, ``{Controlling Queue Delay},'' \emph{Commun. ACM}, vol.~55, no.~7, p. 42–50, jul 2012. [Online]. Available: \url{https://doi.org/10.1145/2209249.2209264}
\BIBentrySTDinterwordspacing

\bibitem{l4srepo}
{L4S development hub}, ``{Linux kernel tree with L4S patches},'' \url{https://github.com/L4STeam/linux}, 2024.

\bibitem{cubic}
\BIBentryALTinterwordspacing
S.~Ha, I.~Rhee, and L.~Xu, ``{CUBIC: a new TCP-friendly high-speed TCP variant},'' \emph{SIGOPS Oper. Syst. Rev.}, vol.~42, no.~5, p. 64–74, jul 2008. [Online]. Available: \url{https://doi.org/10.1145/1400097.1400105}
\BIBentrySTDinterwordspacing

\bibitem{bbrrepo}
Google, ``{BBR - Source code},'' \url{https://github.com/google/bbr}, 2024.

\bibitem{lim-ecn-traversal}
H.~Lim, S.~Kim, J.~Sippe, J.~Kim, G.~White, C.-H. Lee, E.~Wustrow, K.~Lee, D.~Grunwald, and S.~Ha, ``{A Fresh Look at ECN Traversal in the Wild},'' 2022.

\bibitem{bbr-2019-ware}
\BIBentryALTinterwordspacing
R.~Ware, M.~K. Mukerjee, S.~Seshan, and J.~Sherry, ``{Modeling BBR's Interactions with Loss-Based Congestion Control},'' in \emph{{Proceedings of the Internet Measurement Conference}}, ser. IMC '19.\hskip 1em plus 0.5em minus 0.4em\relax New York, NY, USA: Association for Computing Machinery, 2019, p. 137–143. [Online]. Available: \url{https://doi.org/10.1145/3355369.3355604}
\BIBentrySTDinterwordspacing

\bibitem{arun2022starvation}
\BIBentryALTinterwordspacing
V.~Arun, M.~Alizadeh, and H.~Balakrishnan, ``{Starvation in end-to-end congestion control},'' in \emph{{Proceedings of the ACM SIGCOMM 2022 Conference}}, ser. SIGCOMM '22.\hskip 1em plus 0.5em minus 0.4em\relax New York, NY, USA: Association for Computing Machinery, 2022, p. 177–192. [Online]. Available: \url{https://doi.org/10.1145/3544216.3544223}
\BIBentrySTDinterwordspacing

\bibitem{ware-jain-fairness}
\BIBentryALTinterwordspacing
R.~Ware, M.~K. Mukerjee, S.~Seshan, and J.~Sherry, ``{Beyond Jain's Fairness Index: Setting the Bar For The Deployment of Congestion Control Algorithms},'' in \emph{{Proceedings of the 18th ACM Workshop on Hot Topics in Networks}}, ser. HotNets '19.\hskip 1em plus 0.5em minus 0.4em\relax New York, NY, USA: Association for Computing Machinery, 2019, p. 17–24. [Online]. Available: \url{https://doi.org/10.1145/3365609.3365855}
\BIBentrySTDinterwordspacing

\end{thebibliography}

\end{document}